\documentclass[11pt,letterpaper]{article}

\usepackage{libertine}
\usepackage{fullpage}

\usepackage{breakcites}

\usepackage{amsfonts}
\usepackage{amsmath}
\usepackage{amssymb}
\usepackage{amsthm}
\usepackage{float}

\usepackage{cmap}
\usepackage{hyperref}
\usepackage[svgnames]{xcolor}
\hypersetup{colorlinks={true},urlcolor={blue},linkcolor={DarkBlue},citecolor=[named]{DarkGreen}}
\usepackage[authoryear,square]{natbib}

\usepackage{xspace}
\usepackage{fancyhdr}
\usepackage{tcolorbox}

\usepackage{microtype}
\usepackage[capitalise,nameinlink]{cleveref}
\usepackage{enumitem}

\usepackage{doi}

\usepackage{tikz}  %needed for \textcolor as well as tikz
\usetikzlibrary{math}
\usetikzlibrary{arrows}
\usetikzlibrary{patterns,snakes}
\usetikzlibrary{decorations.shapes}
\tikzstyle{overbrace text style}=[font=\tiny, above, pos=.5, yshift=5pt]
\tikzstyle{overbrace style}=[decorate,decoration={brace,raise=5pt,amplitude=3pt}]
\usetikzlibrary{shapes.geometric}

\usepackage{authblk}
\usepackage{xspace}

\allowdisplaybreaks

\newtheorem*{claim*}{Claim}
\newtheorem*{theorem*}{Theorem}
\newtheorem{open}{Open problem}

\theoremstyle{definition}

\newtheorem{example}{Example}
\newtheorem*{votrule}{Voting Rule}

\newcommand{\vv}{\mathbf{v}}

\title{\bf Distortion in Social Choice Problems: \\ The First 15 Years and Beyond}

\author[1]{Elliot Anshelevich}
\author[2]{Aris Filos-Ratsikas}
\author[3]{Nisarg Shah}
\author[4]{Alexandros A. Voudouris}

\affil[1]{Computer Science Department, Rensselaer Polytechnic Institute}
\affil[2]{Department of Computer Science, University of Liverpool}
\affil[3]{Department of Computer Science, University of Toronto}
\affil[4]{School of Computer Science and Electronic Engineering, University of Essex}

\date{}

\begin{document}

\maketitle

\begin{abstract}
The notion of \emph{distortion} in social choice problems has been defined to measure the loss in efficiency---typically measured by the utilitarian social welfare, the sum of utilities of the participating agents---due to having access only to limited information about the preferences of the agents. We survey the most significant results of the literature on distortion from the past 15 years, and highlight important open problems and the most promising avenues of ongoing and future work.
\end{abstract}

\section{Introduction}\label{sec:intro}
Social choice theory \citep{sen1986social} 
is concerned with aggregating the preferences of individuals into a desirable collective decision, and has many applications such as choosing an electoral candidate, a public policy, the recipient of an award, or something as simple as choosing the most appropriate time for a meeting. Founded in the classic theory of \citet{morgenstern1953theory}, these preferences are typically assumed to be captured by \emph{utility functions}, which assign numerical values to the different options, indicating the intensity by which an individual prefers one possible outcome to another. While the existence of such a \emph{cardinal} utility structure is rarely disputed, the predominant approach in social choice theory is to elicit more limited preference information from the participants; in particular, they are usually required to provide \emph{ordinal} preference rankings over the different outcomes. This is primarily due to cognitive reasons, as it is much more conceivable to come up with a ranking based on comparisons rather than a numerical utility structure.  

The inevitable loss of information due to the restricted expressiveness of the preferences makes it rather challenging to optimize objectives of a cardinal nature. A natural such objective is the maximization of the \emph{(utilitarian) social welfare}, defined as the sum of the individual utilities for the chosen outcome. How can we make the right choice when we do not have access to the utilities themselves? As the following example demonstrates, there are cases in which we cannot.   

\begin{example}\label{example:illustrative}
Consider a simple scenario, with two outcomes $a$ and $b$. The ordinal preferences of the participants are such that half of them prefer $a$ to $b$, and the remaining half prefer $b$ to $a$. In such a case, based only on these ordinal preferences, it is impossible to distinguish between the two outcomes, and thus either of them could be selected. So, let us us assume that the winner is $a$. The unknown underlying utilities of the participants for the outcomes might be such that those that prefer $a$ to $b$ are mostly \emph{indifferent} between the two, and have utility about $1/2$ for each of them, whereas the remaining participants \emph{strongly prefer} $b$, and have utility $1$ for $b$ and $0$ for $a$. Consequently, the total utility of the participants for the winner $a$ is only about $n/4$ and their utility for $b$ is $3n/4$, where $n$ is the number of participants. This shows that $a$ is clearly not the optimal choice, and in fact only achieves a social welfare that is a multiplicative factor of $3$ away from the optimal. %\hfill $\qed$
\end{example}

While Example~\ref{example:illustrative} makes it obvious that selecting the outcome that maximizes the social welfare is not always possible, it does not conclusively answer the question of {\em how far away from the optimal outcome our choice can be}. Driven by the principles of worst-case analysis and approximation algorithms~\citep{vazirani2013approximation},
a large part of the computational social choice literature \citep{brandt2016handbook} has studied questions about computing the best outcome for many fundamental social choice problems. In particular, in 2006 \citeauthor{procaccia2006distortion} defined the notion of the \emph{distortion} to measure the worst-case deterioration of an aggregate cardinal objective, such as utilitarian social welfare, in social choice settings, due to having access to preferences of limited expressiveness, particularly ordinal rankings. The associated research agenda posed the following fundamental question:

\begin{quote}
\emph{Which social choice functions achieve the best possible distortion for a given cardinal objective?} 
\end{quote}

Over the past 15 years, this type of question has been studied extensively in the context of many different social choice problems, such as {\em general single-winner elections}, {\em elections with metric preferences}, {\em committee selection}, {\em participatory budgeting}, and {\em matching}, thus giving rise to a rich and vibrant literature at the intersection of economics, computation, and artificial intelligence. In this survey, we highlight the most significant of these results, some current open problems, as well as the most prominent ongoing and future directions.  

\section{Normalized Social Choice}\label{sec:normalized}
The first works on the distortion of social choice rules following \citet{procaccia2006distortion} studied the setting of single-winner elections with normalized utilities. In this setting, there is a set $N$ of $n$ {\em agents} and a set $A$ of $m$ {\em alternatives}. Every agent $i \in N$ has a {\em value} $v_{ij} \in \mathbb{R}_{\geq 0}$ for every alternative $j \in A$; let $\vv = (v_{ij})_{i \in N, j \in A}$. These values are usually assumed to be {\em unit-sum} normalized so that $\sum_{j \in A}v_{ij} = 1$ for every agent $i\in N$. We refer the reader to the note of \citet{aziz2020justifications} for justifications of such normalization assumptions. Given these values, the goal is to compute the alternative that maximizes the {\em social welfare}, which for any alternative $j \in A$ is defined as the total value of the agents for $j$, i.e., $\text{SW}(j | \vv) = \sum_{i \in N} v_{ij}$. It is important to note that while this setting is referred to as an ``election'', and the associated terminology is used, it can broadly capture most decision-making scenarios, where a possible outcome corresponds to an alternative and a participant corresponds to an agent.

If we had access to the values of the agents, maximizing the social welfare would be an easy task; we could simply compute the social welfare of every alternative and then select the one with the largest social welfare. However, the values are the agents' {\em private information}; they are unknown to the designer and it might be the case that they are even unknown to the agents themselves. This stems from the motivation presented in the Introduction, namely that coming up with these values is a cognitively demanding task, which the agents are not often asked to partake in. Instead, we have access to the linear orderings defined by the values. Specifically, every agent $i \in N$ is associated with a (strict) {\em ranking} $\succ_i$ over the alternatives such that $j \succ_i j'$ implies that $v_{ij} \geq v_{ij'}$; let $\succ_\vv = (\succ_i)_{i \in N}$. These rankings are given as input to a {\em social choice function} (or {\em voting rule}) $f$, which decides a single winning alternative $f(\succ_\vv) \in A$. The {\em distortion} of $f$ measures how far away the alternative $f(\succ_\vv)$ is from the optimal alternative in the worst case. Formally, it is defined as the worst-case ratio (over all possible instances) of the maximum social welfare over the social welfare of the alternative returned by $f$, i.e., 
\begin{align*}
\text{distortion}(f) = \sup_{(N,A,\vv)} \frac{\max_{j \in A} \text{SW}(j | \vv)}{\text{SW}( f(\succ_\vv) | \vv)}. 
\end{align*}

\noindent 
For this setting, \citet{procaccia2006distortion} showed that the distortion of the well-known Borda voting rule is unbounded. Another well-known and in fact simpler voting rule is the following.

\begin{votrule}[Plurality]
Elect the alternative that the most agents consider their favorite one, breaking ties arbitrarily.
\end{votrule}

\noindent \citet{caragiannis2011embedding} showed that the Plurality rule has a distortion of $O(m^2)$; this bound was later proven to be best possible among all deterministic voting rules~\citep{caragiannis2017subset}. To give the reader a soft introduction to the type of arguments used to bound the distortion of voting rules, we provide a sketch of the rather simple arguments used for Plurality. We remark that proving bounds on the distortion may require rather involved arguments in general depending on the setting and the type of rules considered. 

\begin{theorem*}\label{claim:plurality}
The distortion of Plurality is $O(m^2)$, and this is best possible among deterministic voting rules.
\end{theorem*}

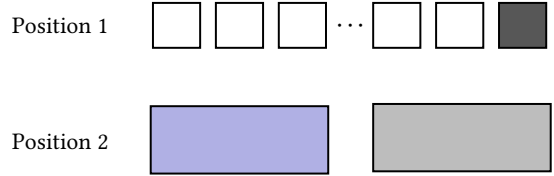
\begin{figure}
    \centering
\tikzset{every picture/.style={line width=0.75pt}} %set default line width to 0.75pt        

\begin{tikzpicture}[x=0.85pt,y=0.85pt,yscale=-1,xscale=1]
%uncomment if require: \path (0,264); %set diagram left start at 0, and has height of 264

%Shape: Rectangle [id:dp6919712843652237] 
\draw  [fill={rgb, 255:red, 0; green, 0; blue, 0 }  ,fill opacity=0.66 ] (509,73) -- (509,93) -- (488,93) -- (488,73) -- cycle ;
%Shape: Rectangle [id:dp5934021295188692] 
\draw   (481,73) -- (481,93) -- (460,93) -- (460,73) -- cycle ;
%Shape: Rectangle [id:dp40839771100018996] 
\draw   (453,73) -- (453,93) -- (432,93) -- (432,73) -- cycle ;
%Shape: Rectangle [id:dp9770635390984419] 
\draw   (411,73) -- (411,93) -- (390,93) -- (390,73) -- cycle ;
%Shape: Rectangle [id:dp4454464766396897] 
\draw   (383,73) -- (383,93) -- (362,93) -- (362,73) -- cycle ;
%Shape: Rectangle [id:dp13163371831418091] 
\draw   (355,73) -- (355,93) -- (334,93) -- (334,73) -- cycle ;
%Shape: Rectangle [id:dp710295069234608] 
\draw  [fill={rgb, 255:red, 155; green, 155; blue, 155 }  ,fill opacity=0.66 ] (511,118) -- (511,148) -- (432,148) -- (432,118) -- cycle ;
%Shape: Rectangle [id:dp6525432110829774] 
\draw  [fill={rgb, 255:red, 150; green, 150; blue, 220 }  ,fill opacity=0.75 ] (412,119) -- (412,149) -- (333,149) -- (333,119) -- cycle ;
% Text Node
\draw (435.6,80) node [anchor=north west][inner sep=0.75pt]  [rotate=-90]  {$\vdots $};
% Text Node
\draw (270,78) node [anchor=north west][inner sep=0.75pt]   [align=left] {{\footnotesize Position 1}};
% Text Node
\draw (270,130) node [anchor=north west][inner sep=0.75pt]   [align=left] {{\footnotesize Position 2}};
\end{tikzpicture}
    \caption{The instance used in the lower bound proof of the theorem.}
    \label{fig:plurality}
\end{figure}

\begin{proof}[Proof Sketch]
For the upper bound, observe that the winning alternative $a$ under Plurality must appear at the first position of the rankings of least $n/m$ agents by the pigeonhole principle. Since the values are unit-sum, these agents must have value at least $1/m$ for $a$, and thus the social welfare of $a$ is at least $n/m^2$. On the other hand, the maximum social welfare of any alternative is $n$, and the bound follows. 

For the lower bound, see the instance of Figure~\ref{fig:plurality}, where (a) there is a set of $m-2$ alternatives such that each of them appears at position 1 in the rankings of $n/(m-2)$ agents, and (b) each of the remaining two alternatives appear at position $2$ in the rankings of $n/2$ agents. Any deterministic voting rule must select either one of the first $m-2$ alternatives or one of the remaining two alternatives, and has no way of distinguishing between alternatives in each group; therefore we can assume without loss of generality that the rule selects either the alternative shaded dark gray or the alternative shaded light gray in the figure. Let the values of the agents that rank the dark gray alternative first be $1/m$ for all alternatives. Let the values of all the remaining agents that rank the light gray alternative second be $1$ for their top-ranked alternative and $0$ for all others. Finally, let the values of all the agents that rank the blue alternative second be $1/2$ for their top two alternatives and zero for the rest. Observe that these values are unit-sum. The social welfare of both the light gray and the dark gray alternatives is $n/(m(m-2))$, whereas the social welfare of the blue alternative is $n/4$, yielding a lower bound $\Omega(m^2)$.
\end{proof}

Randomization proved to be a powerful tool in the quest to achieve improved distortion bounds, overcoming the sharp impossibility results for deterministic rules. A randomized rule outputs a probability distribution over alternatives rather than a single alternative, and its distortion is defined with respect to the expected social welfare that it achieves. The distortion of randomized rules was most notably studied by \citet{boutilier2015optimal}, who designed a particularly involved voting rule with distortion $O(\sqrt{m}\cdot \log^*{m})$. They also presented the following much simpler rule that achieves a slightly inferior bound of $O(\sqrt{m \cdot \log{m}})$: 

\begin{votrule}[\citet{boutilier2015optimal}]
With probability $1/2$, select an alternative uniformly at random, and with probability $1/2$, select every alternative $j \in A$ with probability that is proportional to $j$'s harmonic score, computed based on the harmonic scoring vector $(1, 1/2, ..., 1/m)$, which assigns to every alternative a score of $1/\ell$ for each appearance at position $\ell$ in the rankings of the agents. 
\end{votrule}

\citet{boutilier2015optimal} also proved a general lower bound of $\Omega(\sqrt{m})$ on the achievable distortion of randomized rules, thus effectively settling the problem for the single-winner case. Despite this fact, as we will see in Section~\ref{sec:beyond}, the single-winner normalized social choice setting still offers rather fruitful ground for further work on interesting variants of the problem.

\subsection*{Beyond the single-winner setting}
Many interesting multi-winner extensions have also been studied over the years. \citet{caragiannis2017subset} considered the problem of selecting a subset (committee) of $k$ alternatives that maximizes the social welfare, when the value of each agent for a committee of alternatives is defined as the maximum value she derives from the committee's members. They showed almost tight bounds for both deterministic and randomized rules as functions of $m$ and $k$. \citet{benade2017participatory} considered the participatory budgeting problem, where we are given a budget, each alternative has an associated cost, and the goal is to choose a subset of alternatives so as to maximize the social welfare of the agents while ensuring that the total cost of the chosen alternatives does not exceed the budget. They showed several tight bounds for many different types of input given by the agents, such as knapsack votes, rankings by value, rankings by value for money, and threshold approval votes. Finally, \citet{benade2019rankings} studied the distortion of voting rules that output rankings of alternatives rather than subsets, and showed that bounds that are qualitatively similar to those for the single-winner setting can be achieved by randomized rules.  

\section{Metric Social Choice}\label{sec:metric}
In the normalized social choice setting discussed in the previous section, the values of the agents for the alternatives sum up to $1$ but can otherwise be highly arbitrary. However, in many applications (such as elections), voters usually prefer candidates that express opinions which are closer to their own personal beliefs, and thus the values are determined by an ``ideological distance''. For example, one can envision ideological axes from ``liberal'' to ``conservative'', or from ``libertarian'' to ``authoritarian'', and the preferences of the agents as points in the space defined by these axes. 

More generally, we can think of the agents and the alternatives as points in a (possibly high-dimensional, e.g., corresponding to multiple political issues or ideological dimensions) metric space, in which the triangle inequality is satisfied (see also \citep{merrill1999unified,enelow1984spatial}). Instead of thinking about values and the social welfare objective, in this setting it is more natural to talk about costs and the analogous social cost objective, which we aim to minimize. In particular, the {\em cost} of an agent for an alternative equals their distance in the space, the agents rank the alternatives in a non-decreasing order of their costs, and the {\em social cost} of an alternative is the total cost of the agents for this alternative. Then, the distortion of a voting rule is defined as the ratio of the social cost of the alternative chosen by the rule over the minimum social cost, when given as input the ordinal preferences of the agents.

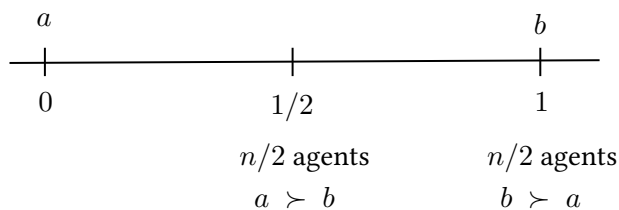
\begin{figure}
    \centering

\tikzset{every picture/.style={line width=0.75pt}} %set default line width to 0.75pt        

\begin{tikzpicture}[x=0.75pt,y=0.75pt,yscale=-0.8,xscale=0.8]
 \path (250,0); %set diagram left start at 0, and has height of 208

%Straight Lines [id:da8406589071234822] 
\draw    (66,65.02) -- (438,63.28) ;
%Straight Lines [id:da8202182479319822] 
\draw    (88.31,55.43) -- (88.31,73.31) ;
%Straight Lines [id:da43890865247703004] 
\draw    (400.63,55.43) -- (400.63,73.31) ;
%Straight Lines [id:da6103076650153427] 
\draw    (244.47,55.43) -- (244.47,73.31) ;

% Text Node
\draw (82.31,81.85) node [anchor=north west][inner sep=0.75pt]    {$0$};
% Text Node
\draw (395.19,81.85) node [anchor=north west][inner sep=0.75pt]    {$1$};
% Text Node
\draw (228.69,81.85) node [anchor=north west][inner sep=0.75pt]    {$1/2$};
% Text Node
\draw (365.19,113.67) node [anchor=north west][inner sep=0.75pt]   [align=left] {$n/2$ agents};
% Text Node
\draw (209.03,113.67) node [anchor=north west][inner sep=0.75pt]   [align=left] {$n/2$ agents};
% Text Node
\draw (81.09,32.12) node [anchor=north west][inner sep=0.75pt]    {$a$};
% Text Node
\draw (394.63,32.12) node [anchor=north west][inner sep=0.75pt]    {$b$};
% Text Node
\draw (373.41,141.18) node [anchor=north west][inner sep=0.75pt]    {$b\ \succ \ a$};
% Text Node
\draw (218.37,141.18) node [anchor=north west][inner sep=0.75pt]    {$a\ \succ \ b$};
\end{tikzpicture}
    \caption{An example showing that no deterministic voting rule can achieve a distortion better than $3$ in the metric social choice setting. Since $n/2$ agents prefer $a$ to $b$ and $n/2$ agents prefer $b$ to $a$, without loss of generality, any rule may select $a$ as the winner for a social cost of $n/4 + n/2 = 3n/4$. On the other hand, the optimal alternative $b$ has social cost $n/4$.}
    \label{fig:distortion_example}
\end{figure}

An example very similar to Example~\ref{example:illustrative} shows that the distortion of any deterministic voting rule in the metric social choice setting is at least $3$; see Figure~\ref{fig:distortion_example}. However, in contrast to the normalized setting from Section~\ref{sec:normalized}, the metric domain restriction makes it possible to achieve small constant bounds, even with deterministic rules. This was first shown by \citet{anshelevich2015approximating}, who initiated the study of distortion in the metric setting and showed that the well-known Copeland rule~\citep{copeland1951reasonable} achieves a distortion of $5$.

\citet{anshelevich2015approximating} further conjectured that the lower bound of $3$ provided by the example of Figure~\ref{fig:distortion_example} must be the tight bound for deterministic rules, and in fact that it can be achieved by the well-known Ranked Pairs rule~\citep{tideman1987independence}.
In the years that followed, a succession of papers made significant progress towards resolving this conjecture. \citet{skowron2017stv} showed that the distortion of every positional scoring rule is $\Omega(\sqrt{\log{m}})$, without however providing any scoring rule with distortion better than $O(m)$. They also showed that the distortion of the Single Transferable Vote (STV) rule is between $\Omega({\sqrt{\log{m}})}$ and $O(\log{m})$. \citet{goel2017metric} were the first to disprove that the Ranked Pairs rule matches the bound of $3$, by presenting a lower bound of $5$ for it; its exact distortion was later shown to be $\Theta(\sqrt{m})$ by \citet{kempe2020duality}. \citet{munagala2019improved} made significant progress towards resolving the conjecture by designing a novel voting rule that broke the barrier of $5$ and achieved distortion $4.236$. They also identified many important properties that would suffice to achieve the conjectured bound of $3$. Based on these properties, the conjecture was finally positively resolved very recently by \citet{gkatzelis2020resolving}, who designed the following {\em plurality matching} voting rule.

\begin{votrule}[\citet{gkatzelis2020resolving}]
For every alternative $j \in A$, construct the integral domination graph of $j$, which is a bipartite graph with the nodes of both sides corresponding to agents, and each edge $(x,y)$ between agents $x$ and $y$ indicating that $x$ prefers alternative $j$ to the favorite alternative of $y$. Return an alternative whose integral domination graph admits a perfect matching; such an alternative is guaranteed to exist. 
\end{votrule}

These results have impressive and counter-intuitive implications. They state that, as long as agents and alternatives correspond to points in a metric space, it is always possible to choose a close-to-optimal alternative based only on the ordinal rankings of the agents, without even the need to know their actual locations in the space.

\subsection*{Randomized Rules}
Randomized voting rules have also been considered in the metric setting by \citet{anshelevich2017randomized}, who showed that randomization is able to break the barrier of $3$ suggested by the example of Figure~\ref{fig:distortion_example}, albeit only by a small amount which vanishes as the number of agents becomes large. In particular, they showed that {\em Random Dictatorship} (which selects an agent uniformly at random and outputs her favorite alternative) has distortion $3-2/n$, and complemented this with a lower bound of $2$ on the distortion of any randomized rule, which holds even when the metric space is a simple line. \citet{kempe2020communication} presented a randomized rule that has distortion $3-2/m$, which is slightly better than that of Random Dictatorship when $n$ is large but $m$ is small. However, the following is still a major open question. 

\begin{open}
What is the best possible distortion for randomized social choice rules in the metric setting?
\end{open}

Progress has been made for some special cases, such as the line metric, for which \citet{feldman2016voting} presented a randomized rule, the \emph{Spike mechanism}, with a tight distortion bound of $2$. %, matching the lower bound of $2$ shown independently by the authors of that paper and \citet{anshelevich2017randomized}. 
\citet{fain2017sequential} also considered restricted spaces called median graphs, and showed that it is possible to achieve a distortion of almost $1.218$, via a \emph{sequential deliberation} rule. Finally, \citet{anshelevich2017randomized} also introduced the notion of $\alpha$-\emph{decisiveness} to measure the preference strength of an agent for her top choice over her second choice, and showed upper and lower bounds parameterized by this quantity.

\subsection*{Limited Ordinal Information}
While the bulk of papers on distortion in metric social choice settings have focused on rules that in general take as input complete rankings, there has also been work on rules that require incomplete ordinal information, motivated by the fact that some rules like Plurality and Random Dictatorship essentially only need to know the favorite alternatives of the agents. In this context, \citet{gross2017vote} showed a lower bound of $3-2/\lfloor{m/k\rfloor}$ on the distortion when the agents submit top-$k$ rankings for $k < m/2$, and showed that a very simple randomized rule, which asks random agents for their favorite alternatives until two of them agree, has many nice axiomatic properties and is almost optimal in terms of distortion when the number of alternatives is not too large. 

\citet{fain2019random} studied voting rules that require the agents to compare only a constant number of pairs of alternatives (constant sample complexity) and presented the {\em Random Referee} rule, which achieves a constant {\em squared distortion}, a measure of the variance of the distortion of randomized rules. Furthermore, for any constant $k$, they showed that no rule that elicits only top-$k$ rankings can have squared distortion that is sublinear in $m$, and presented the {\em Random Oligarchy} rule with distortion almost matching the lower bound of $3-2/m$ for $k=1$. As mentioned above, this small gap was later resolved by \citet{kempe2020communication}. He mainly focused on deterministic rules and showed that the distortion of rules that take as input the alternatives that the agents rank in a given set of $k$ positions is $\Theta(m/k)$. Moreover, he showed that the distortion of rules that require the agents to submit $b$ bits of information about their rankings is $\Omega(m/b)$.

\subsection*{Other Settings} 
Several papers have also studied variations of the main metric setting by making different assumptions about the origin of the alternatives, the behavior of the agents, and the objective. In particular, \citet{Cheng2017representative} considered a probabilistic model in which there are only two alternatives that are independently drawn from the population of agents. They showed that the distortion of the majority rule is between $1.5$ and $2$ for arbitrary metric spaces, and improves to $1.716$ for the line metric. In follow-up work, \citet{cheng2018multiple} extended their analysis to the case where multiple alternatives are randomly chosen from the set of agents, and showed bounds on the distortion of positional scoring rules. \citet{borodin2019primaries} studied {\em primary voting systems} wherein there are political parties and agents affiliated with those vote to choose the party representatives, who then compete in the general elections. Their main results were that every voting rule is guaranteed to perform almost as well under the primary system as under the direct system (in which all alternatives are candidates in the election), but not vice versa.

\citet{ghodsi2019abstention} considered settings with two alternatives, in which some agents are allowed to abstain from the election process in case they are pretty much indifferent between the alternatives or the distance from their favorite alternative is sufficiently large. They presented a probabilistic model to capture these properties and characterized the distortion of the majority rule.

\citet{pierczynski2019approval} focused on approval-based preferences according to which the agents do not rank the alternatives, but instead list those that they approve. Besides showing that the distortion of the approval rule is unbounded, they also introduced the notion of {\em acceptability-based distortion}, which aims to capture how far away the chosen alternative is from the alternative that is approved by the most agents. For this definition of distortion, they showed bounds for well-known rules such as Plurality, Borda, Ranked Pairs, and Copeland. \citet{jaworski20evaluating} studied the distortion of multi-winner rules for the election of a committee, whose quality is determined by the extent to which the decisions it makes on binary issues are consistent to the preferences of the agents on those issues.

\section{Other Problems and Extensions}\label{sec:extensions}
Distortion has also been considered in many other settings that have either been studied under the general umbrella of social choice theory, or are very much related.

\subsection*{Matching and Other Graph Problems}
The \emph{one-sided matching} problem (or {\em house allocation}) is a classic problem in economics \citep{hylland1979efficient}, where $n$ agents have preferences over a set of $n$ items, %a set of agents has preferences over a set of items, 
and the goal is to match each agent to a single item in order to maximize the social welfare, defined as the total value of the agents for the items they are matched to. The problem was first studied in the context of distortion by \citet{filos-Ratsikas2014matching}, who showed that \emph{Random Serial Dictatorship} (e.g., see \citep{abdulkadirouglu1998random}) has distortion $O(\sqrt{n})$, and furthermore, this is best possible among all rules that take as input the ordinal preferences of the agents induced by their underlying normalized values. Recently, for the same setting, \citet{amanatidis2021matching} showed that the best possible deterministic rule for one-sided matching has distortion $\Theta(n^2)$.

\citeauthor{anshelevich2016blind}~[\citeyear{anshelevich2016blind,anshelevich2016truthful}] and \citet{abramowitz2017utilitarians} studied the distortion in {\em maximum weight matching} and many other graph problems such as {\em maximum $k$-Sum}, {\em densest $k$-Subgraph}, and {\em maximum travelling salesperson (TSP)}. They primarily focused on the case of symmetric values that satisfy the triangle inequality and presented a general framework for designing greedy rules that achieve small constant distortion bounds. For instance, among other results, they showed that deterministic rules can achieve a distortion bound of $2$ for maximum-weight matching and maximum-TSP, whereas randomized rules can achieve bounds of $1.6$ and $1.88$ for those two problems respectively. \citet{anshelevich2018ordinal} studied a general facility assignment model with the goal of coming up with a (possibly constrained) assignment of agents to facilities. For many different optimization objectives, they showed a black-box reduction which converts a cardinal rule for the problem into an ordinal rule with small distortion. In particular, they showed that the distortion of many facility assignment and matching problems is at most $3$ when the possible locations of the facilities are public knowledge, but only ordinal preferences of the agents for the facilities are known. 

\subsection*{Truthfulness and Distortion}
In most of the aforementioned results, the implicit assumption was that when agents are asked to submit their ordinal preference rankings, they do so honestly. In many cases, however, the agents might have incentives to misreport their preferences, if that could result in a more favorable outcome (e.g., the election of a more preferred alternative, or the allocation of a more preferable item). In such a case, we would like to design rules that do not provide incentives for such strategic behavior; these are typically called \emph{truthful mechanisms} and are rooted in the principles of \emph{mechanism design} (e.g, see \citep{nisan2007introduction}). 

For the normalized social choice setting, \citet{bhaskar2018truthful} studied multi-winner elections and participatory budgeting problems. The main takeaway from their results was that essentially the best randomized truthful mechanisms are ordinal, and are in fact asymptotically almost as good (up to logarithmic factors) as the best ordinal non-truthful rules. The single-winner setting was studied previously by \citet{filos2014truthful}, albeit for a different normalization (known as {\em unit-range}), and later on by \citet{caragiannis2018truthful} for a related hybrid social choice setting. In the metric setting, and the facility location problem in particular, \citet{feldman2016voting} considered different degrees of expressiveness in terms of the agents' preferences, and proved upper and lower bounds on the distortion of truthful mechanisms. 

Even before \citet{bhaskar2018truthful} reached a similar conclusion, \citet{filos-Ratsikas2014matching} observed that the best truthful mechanisms are ordinal in the one-sided matching setting described earlier. In particular, they showed that the $O(\sqrt{n})$ bound of random serial dictatorship (which is both ordinal and truthful) is best possible among all truthful mechanisms. \citet{caragiannis2016truthful} studied the same rule for the minimum-cost metric matching problem, and showed that its distortion is between $\Omega(n^{0.29})$ and $O(n)$. This is in fact the best known upper bound for any ordinal rule in this setting, even non-truthful ones. To this end, we have the following open problem.

\begin{open}
What is the distortion of the best ordinal rule for the minimum-cost metric matching problem? What if we also require truthfulness?
\end{open}

\section{Generalizations of Distortion}\label{sec:beyond}
While there are still several interesting open questions in the classic settings described above, in this section we would like to highlight a more recent stream of work which views the distortion as a more general concept of measuring the loss of efficiency due to having access to incomplete information. We believe this opens up many intriguing avenues for future investigations. In particular, these works refine the notion of distortion to capture scenarios in which the loss of efficiency is not necessarily because the agents express ordinal preferences, but because there is only \emph{limited access} to the {\em complete} cardinal information. Some works consider settings in which the provided information is in a sense ``less than cardinal but more than ordinal'', whereas others explore the limits of communication between the agents and the designer, which inherently results in some information loss. Finally, some recent papers consider the loss of information that is incurred by making decisions in a distributed manner. We highlight some of the recent results below, as well as areas for future work. 

\subsection*{Beyond Ordinal Information}
As discussed in the Introduction, the standard argument in favor of ordinal preferences is based on the cognitive limitations of the agents. However, one can conceivably consider settings where some \emph{limited additional} information of a cardinal nature is available, without imposing a large cognitive overhead on the agents.  

Along those lines, \citet{amanatidis2020peeking} studied deterministic rules for single-winner social choice that have access to the ordinal preferences of the agents and can also learn some of their numerical values by appropriately asking a small number of \emph{queries} to the agents. They showed that a single query per agent is enough to achieve distortion $O(m)$ (improving the bound of $\Theta(m^2)$ achieved by ordinal rules), and designed a rule that (as special cases) achieves distortion $O(\sqrt{m})$ by making $O(\log{m})$ queries per agent (thus matching the best bound possible for ordinal \emph{randomized} rules), and \emph{constant} distortion by making $O(\log^2{m})$ queries per agent; quite remarkably, these bounds holds even without any normalization assumptions.
Moreover, the authors showed that it is impossible to obtain sublinear distortion with a single query per agent, and also that $\omega(\log{m})$ queries per agent are necessary to achieve constant distortion.

\citet{amanatidis2020peeking} conjectured that it should be possible to achieve distortion $O(\sqrt{m})$ by asking only a constant number of queries per agent and constant distortion by asking $\Theta(\log{m})$ queries per agent. Another interesting direction is to consider randomized rules in the context of this setting, where randomization can be used in either the query phase, the election phase, or both. This line of work raises the following conceptual open question.

\begin{open}
Which other settings admit interesting tradeoffs between the distortion and the number of queries?
\end{open}

Follow-up works \citep{amanatidis2021matching,ma2020matching} identified the one-sided matching problem and several of its generalizations as such settings, and showed quantitatively similar tradeoffs between the number of queries and the distortion. It would be very interesting to see if a similar approach can be applied to the metric social choice setting, where the bounds are already small constants and significant improvements are more difficult to obtain.

In the metric setting, \citet{abramowitz2019awareness} also considered the case where the agents provide limited cardinal information on top of their ordinal preferences, by indicating preference strengths (the voters' \emph{passion}). In particular, for each agent $i$ and every pair of alternatives $a$ and $b$, the voting rule has access to the ratio of the distances of agent $i$ to alternatives $a$ and $b$. The authors showed that there exist deterministic rules with distortion $\sqrt{2}$ when there only two alternatives, and distortion $2$ when there are more alternatives. They also considered rules that do not know the exact preference strengths, but only whether these strengths exceed a predefined threshold $\tau$, and showed how the distortion bounds change as functions of $\tau$. Perhaps the most general conceptual research direction is the following.

\begin{open}
What other meaningful ways of eliciting limited cardinal information can be used to improve the distortion in metric or normalized social choice?
\end{open}

We remark that the idea of eliciting cardinal information has also been present in earlier works. In particular, \citet{benade2017participatory}, and later on \citet{bhaskar2018truthful}, considered (deterministic or randomized) \emph{threshold approval} queries, in which the agents are asked to list all the alternatives that they value higher than a specified threshold. 

\subsection*{Communication complexity}
\citet{mandal2019thrifty} considered similar tradeoffs between information and distortion, but without assuming a priori access to the ordinal preferences of the agents. Instead, they focused on voting rules that consist of an elicitation part which collects a specific number of bits of information per agent about their values, and an aggregation part which uses this information to decide a single winning alternative. \citet{mandal2019thrifty} proposed a voting rule that uses deterministic elicitation and aggregation, and achieves distortion $d$ with communication complexity $O(m \log(d\log{m})/d)$. Furthermore, they showed that to achieve distortion $d$, any voting rule using deterministic elicitation must have communication complexity $\Omega(m/d^2)$, whereas any voting rule using randomized elicitation must have communication complexity $\Omega(m/d^3)$. 

In follow-up work, \citet{mandal2020optimal} improved these tradeoffs by showing tight bounds (up to logarithmic factors) on the communication complexity necessary to achieve distortion $d$. Specifically, by exploiting recent advances in streaming algorithms and making a connection to the literature on communication complexity from theoretical computer science, they showed a bound of $\tilde{\Theta}(m/d)$ for deterministic elicitation and a bound of $\tilde{\Theta}(m/d^3)$ for randomized elicitation. \citet{mandal2020optimal} also considered the multi-winner setting of  \citet{caragiannis2017subset}, and showed that the required communication complexity for achieving distortion $d$ is $\tilde{\Theta}(m/(kd))$ for deterministic elicitation and $\tilde{\Theta}(m/(kd^3))$ for randomized elicitation, where $k$ is the number of winners selected. % for which they again showed tight bounds on the communication complexity necessary to achieve a given distortion bound. 
While these two papers have provided an almost complete picture of the communication complexity of single- and multi-winner voting rules, one could ask similar questions for other settings.  

\begin{open}
Which other settings admit interesting tradeoffs between distortion and communication? 
\end{open}
Settings like the one-sided matching and its generalizations seem like good candidates, and so does the metric social choice setting.

\subsection*{Distributed settings}
In all the papers discussed so far, the voting rules operate by collecting the preferences of all agents and then aggregating them into a common decision in a single step. However, in many important applications (such as presidential elections in the US) the aggregation process is {\em distributed}, in the sense that the agents are partitioned into {\em districts} and vote locally therein to choose representative alternatives, which are then aggregated into a final collective outcome. With this in mind, \citet{FMV2020distributed} initiated the study of the distortion in distributed settings, aiming to capture the loss of efficiency not only due to the possibly limited expressiveness of the agents' preferences, but also due to not having full access to the details of the local decisions within the districts. 

\citet{FMV2020distributed} focused on the unit-sum single-winner setting and bounded the distortion of deterministic distributed rules, which use classic voting rules for the local district elections, and choose as overall winner the alternative that is the representative of the most districts (or has the largest total weight among the representatives in case the districts are weighted). Their results showed that the distributed nature of such mechanisms leads to considerably larger distortion compared to their centralized counterparts. Later on, \citet{FV2020facility} studied the distortion of deterministic distributed rules for discrete and continuous facility location on the line (single-dimensional metric setting), and showed tight bounds for the social cost objective that are small constants, both in general and under truthfulness constraints. There are many interesting open questions to be answered in distributed settings. 

\begin{open}
What is the distortion of randomized distributed rules? What happens when the goal is to choose a committee or locate more facilities? What about general metric spaces and other settings?
\end{open}

\bibliographystyle{plainnat}
\bibliography{distortion}

\begin{thebibliography}{53}
\providecommand{\natexlab}[1]{#1}
\providecommand{\url}[1]{\texttt{#1}}
\expandafter\ifx\csname urlstyle\endcsname\relax
  \providecommand{\doi}[1]{doi: #1}\else
  \providecommand{\doi}{doi: \begingroup \urlstyle{rm}\Url}\fi

\bibitem[Abdulkadiro{\u{g}}lu and S{\"o}nmez(1998)]{abdulkadirouglu1998random}
Atila Abdulkadiro{\u{g}}lu and Tayfun S{\"o}nmez.
\newblock Random serial dictatorship and the core from random endowments in
  house allocation problems.
\newblock \emph{Econometrica}, 66\penalty0 (3):\penalty0 689--701, 1998.

\bibitem[Abramowitz and Anshelevich(2018)]{abramowitz2017utilitarians}
Ben Abramowitz and Elliot Anshelevich.
\newblock Utilitarians without utilities: Maximizing social welfare for graph
  problems using only ordinal preferences.
\newblock In \emph{Proceedings of the 32nd {AAAI} Conference on Artificial
  Intelligence ({AAAI})}, pages 894--901, 2018.

\bibitem[Abramowitz et~al.(2019)Abramowitz, Anshelevich, and
  Zhu]{abramowitz2019awareness}
Ben Abramowitz, Elliot Anshelevich, and Wennan Zhu.
\newblock Awareness of voter passion greatly improves the distortion of metric
  social choice.
\newblock In \emph{Proceedings of the The 15th Conference on Web and Internet
  Economics (WINE)}, pages 3--16, 2019.

\bibitem[Amanatidis et~al.(2020)Amanatidis, Birmpas, Filos-Ratsikas, and
  Voudouris]{amanatidis2020peeking}
Georgios Amanatidis, Georgios Birmpas, Aris Filos-Ratsikas, and Alexandros~A.
  Voudouris.
\newblock Peeking behind the ordinal curtain: Improving distortion via cardinal
  queries.
\newblock In \emph{Proceedings of the 34th {AAAI} Conference on Artificial
  Intelligence ({AAAI})}, pages 1782--1789, 2020.

\bibitem[Amanatidis et~al.(2021)Amanatidis, Birmpas, Filos-Ratsikas, and
  Voudouris]{amanatidis2021matching}
Georgios Amanatidis, Georgios Birmpas, Aris Filos-Ratsikas, and Alexandros~A.
  Voudouris.
\newblock A few queries go a long way: Information-distortion tradeoffs in
  matching.
\newblock In \emph{Proceedings of the 35th {AAAI} Conference on Artificial
  Intelligence ({AAAI})}, page forthcoming, 2021.

\bibitem[Anshelevich and Postl(2017)]{anshelevich2017randomized}
Elliot Anshelevich and John Postl.
\newblock Randomized social choice functions under metric preferences.
\newblock \emph{Journal of Artificial Intelligence Research}, 58:\penalty0
  797--827, 2017.

\bibitem[Anshelevich and Sekar(2016{\natexlab{a}})]{anshelevich2016blind}
Elliot Anshelevich and Shreyas Sekar.
\newblock Blind, greedy, and random: Algorithms for matching and clustering
  using only ordinal information.
\newblock In \emph{Proceedings of the 30th {AAAI} Conference on Artificial
  Intelligence ({AAAI})}, pages 390--396, 2016{\natexlab{a}}.

\bibitem[Anshelevich and Sekar(2016{\natexlab{b}})]{anshelevich2016truthful}
Elliot Anshelevich and Shreyas Sekar.
\newblock Truthful mechanisms for matching and clustering in an ordinal world.
\newblock In \emph{Proceedings of the 12th Conference on Web and Internet
  Economics ({WINE})}, pages 265--278, 2016{\natexlab{b}}.

\bibitem[Anshelevich and Zhu(2018)]{anshelevich2018ordinal}
Elliot Anshelevich and Wennan Zhu.
\newblock Ordinal approximation for social choice, matching, and facility
  location problems given candidate positions.
\newblock In \emph{Proceedings of the 14th International Conference on Web and
  Internet Economics ({WINE})}, pages 3--20, 2018.

\bibitem[Anshelevich et~al.(2015)Anshelevich, Bhardwaj, and
  Postl]{anshelevich2015approximating}
Elliot Anshelevich, Onkar Bhardwaj, and John Postl.
\newblock Approximating optimal social choice under metric preferences.
\newblock In \emph{Proceedings of the 29th {AAAI} Conference on Artificial
  Intelligence ({AAAI})}, pages 777--783, 2015.

\bibitem[Aziz(2020)]{aziz2020justifications}
Haris Aziz.
\newblock Justifications of welfare guarantees under normalized utilities.
\newblock \emph{ACM SIGecom Exchanges}, 17\penalty0 (2):\penalty0 71--75, 2020.

\bibitem[Benad{\`{e}} et~al.(2017)Benad{\`{e}}, Nath, Procaccia, and
  Shah]{benade2017participatory}
Gerdus Benad{\`{e}}, Swaprava Nath, Ariel~D. Procaccia, and Nisarg Shah.
\newblock Preference elicitation for participatory budgeting.
\newblock In \emph{Proceedings of the 31st {AAAI} Conference on Artificial
  Intelligence ({AAAI})}, pages 376--382, 2017.

\bibitem[Benad{\`{e}} et~al.(2019)Benad{\`{e}}, Procaccia, and
  Qiao]{benade2019rankings}
Gerdus Benad{\`{e}}, Ariel~D. Procaccia, and Mingda Qiao.
\newblock Low-distortion social welfare functions.
\newblock In \emph{Proceedings of the 33rd {AAAI} Conference on Artificial
  Intelligence ({AAAI})}, pages 1788--1795, 2019.

\bibitem[Bhaskar et~al.(2018)Bhaskar, Dani, and Ghosh]{bhaskar2018truthful}
Umang Bhaskar, Varsha Dani, and Abheek Ghosh.
\newblock Truthful and near-optimal mechanisms for welfare maximization in
  multi-winner elections.
\newblock In \emph{Proceedings of the 32nd {AAAI} Conference on Artificial
  Intelligence ({AAAI})}, pages 925--932, 2018.

\bibitem[Borodin et~al.(2019)Borodin, Lev, Shah, and
  Strangway]{borodin2019primaries}
Allan Borodin, Omer Lev, Nisarg Shah, and Tyrone Strangway.
\newblock Primarily about primaries.
\newblock In \emph{Proceedings of the 33rd {AAAI} Conference on Artificial
  Intelligence ({AAAI})}, pages 1804--1811, 2019.

\bibitem[Boutilier et~al.(2015)Boutilier, Caragiannis, Haber, Lu, Procaccia,
  and Sheffet]{boutilier2015optimal}
Craig Boutilier, Ioannis Caragiannis, Simi Haber, Tyler Lu, Ariel~D. Procaccia,
  and Or~Sheffet.
\newblock Optimal social choice functions: A utilitarian view.
\newblock \emph{Artificial Intelligence}, 227:\penalty0 190--213, 2015.

\bibitem[Brandt et~al.(2016)Brandt, Conitzer, Endriss, Lang, and
  Procaccia]{brandt2016handbook}
Felix Brandt, Vincent Conitzer, Ulle Endriss, J{\'e}r{\^o}me Lang, and Ariel~D
  Procaccia.
\newblock \emph{Handbook of computational social choice}.
\newblock Cambridge University Press, 2016.

\bibitem[Caragiannis and Procaccia(2011)]{caragiannis2011embedding}
Ioannis Caragiannis and Ariel~D. Procaccia.
\newblock Voting almost maximizes social welfare despite limited communication.
\newblock \emph{Artificial Intelligence}, 175\penalty0 (9-10):\penalty0
  1655--1671, 2011.

\bibitem[Caragiannis et~al.(2016)Caragiannis, Filos-Ratsikas, Frederiksen,
  Hansen, and Tan]{caragiannis2016truthful}
Ioannis Caragiannis, Aris Filos-Ratsikas, S{\o}ren Kristoffer~Stiil
  Frederiksen, Kristoffer~Arnsfelt Hansen, and Zihan Tan.
\newblock Truthful facility assignment with resource augmentation: An exact
  analysis of serial dictatorship.
\newblock In \emph{Proceedings of the 12th International Conference on Web and
  Internet Economics ({WINE})}, pages 236--250, 2016.

\bibitem[Caragiannis et~al.(2017)Caragiannis, Nath, Procaccia, and
  Shah]{caragiannis2017subset}
Ioannis Caragiannis, Swaprava Nath, Ariel~D. Procaccia, and Nisarg Shah.
\newblock Subset selection via implicit utilitarian voting.
\newblock \emph{Journal of Artificial Intelligence Research}, 58:\penalty0
  123--152, 2017.

\bibitem[Caragiannis et~al.(2018)Caragiannis, Filos{-}Ratsikas, Nath, and
  Voudouris]{caragiannis2018truthful}
Ioannis Caragiannis, Aris Filos{-}Ratsikas, Swaprava Nath, and Alexandros~A.
  Voudouris.
\newblock Truthful mechanisms for ownership transfer with expert advice.
\newblock \emph{CoRR}, abs/1802.01308, 2018.

\bibitem[Cheng et~al.(2017)Cheng, Dughmi, and Kempe]{Cheng2017representative}
Yu~Cheng, Shaddin Dughmi, and David Kempe.
\newblock Of the people: Voting is more effective with representative
  candidates.
\newblock In \emph{Proceedings of the 2017 {ACM} Conference on Economics and
  Computation ({EC})}, pages 305--322, 2017.

\bibitem[Cheng et~al.(2018)Cheng, Dughmi, and Kempe]{cheng2018multiple}
Yu~Cheng, Shaddin Dughmi, and David Kempe.
\newblock On the distortion of voting with multiple representative candidates.
\newblock In \emph{Proceedings of the 32nd {AAAI} Conference on Artificial
  Intelligence ({AAAI})}, pages 973--980, 2018.

\bibitem[Copeland(1951)]{copeland1951reasonable}
Arthur~H. Copeland.
\newblock A reasonable social welfare function.
\newblock Technical report, mimeo, 1951. University of Michigan, 1951.

\bibitem[Enelow and Hinich(1984)]{enelow1984spatial}
James~M. Enelow and Melvin~J. Hinich.
\newblock \emph{The spatial theory of voting: An introduction}.
\newblock CUP Archive, 1984.

\bibitem[Fain et~al.(2017)Fain, Goel, Munagala, and
  Sakshuwong]{fain2017sequential}
Brandon Fain, Ashish Goel, Kamesh Munagala, and Sukolsak Sakshuwong.
\newblock Sequential deliberation for social choice.
\newblock In \emph{International Conference on Web and Internet Economics},
  pages 177--190, 2017.

\bibitem[Fain et~al.(2019)Fain, Goel, Munagala, and Prabhu]{fain2019random}
Brandon Fain, Ashish Goel, Kamesh Munagala, and Nina Prabhu.
\newblock Random dictators with a random referee: Constant sample complexity
  mechanisms for social choice.
\newblock In \emph{Proceedings of the 33rd {AAAI} Conference on Artificial
  Intelligence ({AAAI})}, pages 1893--1900, 2019.

\bibitem[Feldman et~al.(2016)Feldman, Fiat, and Golomb]{feldman2016voting}
Michal Feldman, Amos Fiat, and Iddan Golomb.
\newblock On voting and facility location.
\newblock In \emph{Proceedings of the 2016 {ACM} Conference on Economics and
  Computation ({EC})}, pages 269--286, 2016.

\bibitem[Filos-Ratsikas and Miltersen(2014)]{filos2014truthful}
Aris Filos-Ratsikas and Peter~Bro Miltersen.
\newblock Truthful approximations to range voting.
\newblock In \emph{Proceedings of the 10th International Conference on Web and
  Internet Economics ({WINE})}, pages 175--188, 2014.

\bibitem[Filos{-}Ratsikas and Voudouris(2020)]{FV2020facility}
Aris Filos{-}Ratsikas and Alexandros~A. Voudouris.
\newblock Approximate mechanism design for distributed facility location.
\newblock \emph{CoRR}, abs/2007.06304, 2020.

\bibitem[Filos{-}Ratsikas et~al.(2014)Filos{-}Ratsikas, Frederiksen, and
  Zhang]{filos-Ratsikas2014matching}
Aris Filos{-}Ratsikas, S{\o}ren Kristoffer~Stiil Frederiksen, and Jie Zhang.
\newblock Social welfare in one-sided matchings: Random priority and beyond.
\newblock In \emph{Proceedings of the 7th Symposium on Algorithmic Game Theory
  ({SAGT})}, pages 1--12, 2014.

\bibitem[Filos-Ratsikas et~al.(2020)Filos-Ratsikas, Micha, and
  Voudouris]{FMV2020distributed}
Aris Filos-Ratsikas, Evi Micha, and Alexandros~A. Voudouris.
\newblock The distortion of distributed voting.
\newblock \emph{Artificial Intelligence}, 286:\penalty0 103343, 2020.

\bibitem[Ghodsi et~al.(2021)Ghodsi, Latifian, and
  Seddighin]{ghodsi2019abstention}
Mohammad Ghodsi, Mohamad Latifian, and Masoud Seddighin.
\newblock On the distortion value of the elections with abstention.
\newblock \emph{Journal of Artificial Intelligence Research}, 70:\penalty0
  567--595, 2021.

\bibitem[Gkatzelis et~al.(2020)Gkatzelis, Halpern, and
  Shah]{gkatzelis2020resolving}
Vasilis Gkatzelis, Daniel Halpern, and Nisarg Shah.
\newblock Resolving the optimal metric distortion conjecture.
\newblock In \emph{Proceedings of the 61st {IEEE} Annual Symposium on
  Foundations of Computer Science ({FOCS})}, pages 1427--1438, 2020.

\bibitem[Goel et~al.(2017)Goel, Krishnaswamy, and Munagala]{goel2017metric}
Ashish Goel, Anilesh~K Krishnaswamy, and Kamesh Munagala.
\newblock Metric distortion of social choice rules: Lower bounds and fairness
  properties.
\newblock In \emph{Proceedings of the 2017 {ACM} Conference on Economics and
  Computation (EC)}, pages 287--304, 2017.

\bibitem[Gross et~al.(2017)Gross, Anshelevich, and Xia]{gross2017vote}
Stephen Gross, Elliot Anshelevich, and Lirong Xia.
\newblock Vote until two of you agree: Mechanisms with small distortion and
  sample complexity.
\newblock In \emph{Proceedings of the 31st {AAAI} Conference on Artificial
  Intelligence ({AAAI})}, pages 544--550, 2017.

\bibitem[Hylland and Zeckhauser(1979)]{hylland1979efficient}
Aanund Hylland and Richard Zeckhauser.
\newblock The efficient allocation of individuals to positions.
\newblock \emph{Journal of Political economy}, 87\penalty0 (2):\penalty0
  293--314, 1979.

\bibitem[Jaworski and Skowron(2020)]{jaworski20evaluating}
Michal Jaworski and Piotr Skowron.
\newblock Evaluating committees for representative democracies: The distortion
  and beyond.
\newblock In \emph{Proceedings of the 29th International Joint Conference on
  Artificial Intelligence ({IJCAI})}, pages 196--202, 2020.

\bibitem[Kempe(2020{\natexlab{a}})]{kempe2020communication}
David Kempe.
\newblock Communication, distortion, and randomness in metric voting.
\newblock In \emph{Proceedings of the 34th {AAAI} Conference on Artificial
  Intelligence ({AAAI})}, pages 2087--2094, 2020{\natexlab{a}}.

\bibitem[Kempe(2020{\natexlab{b}})]{kempe2020duality}
David Kempe.
\newblock An analysis framework for metric voting based on {LP} duality.
\newblock In \emph{Proceedings of the 34th {AAAI} Conference on Artificial
  Intelligence ({AAAI})}, pages 2079--2086, 2020{\natexlab{b}}.

\bibitem[Ma et~al.(2020)Ma, Menon, and Larson]{ma2020matching}
Thomas Ma, Vijay Menon, and Kate Larson.
\newblock Improving welfare in one-sided matching using simple threshold
  queries.
\newblock \emph{CoRR}, abs/2011.13977, 2020.

\bibitem[Mandal et~al.(2019)Mandal, Procaccia, Shah, and
  Woodruff]{mandal2019thrifty}
Debmalya Mandal, Ariel~D. Procaccia, Nisarg Shah, and David~P. Woodruff.
\newblock Efficient and thrifty voting by any means necessary.
\newblock In \emph{Proceedings of the 32nd Annual Conference on Neural
  Information Processing Systems ({NeurIPS})}, pages 7178--7189, 2019.

\bibitem[Mandal et~al.(2020)Mandal, Shah, and Woodruff]{mandal2020optimal}
Debmalya Mandal, Nisarg Shah, and David~P. Woodruff.
\newblock Optimal communication-distortion tradeoff in voting.
\newblock In \emph{Proceedings of the 21st {ACM} Conference on Economics and
  Computation ({EC})}, pages 795--813, 2020.

\bibitem[Merrill~III et~al.(1999)Merrill~III, Merrill, and
  Grofman]{merrill1999unified}
Samuel Merrill~III, Samuel Merrill, and Bernard Grofman.
\newblock \emph{A unified theory of voting: Directional and proximity spatial
  models}.
\newblock Cambridge University Press, 1999.

\bibitem[Munagala and Wang(2019)]{munagala2019improved}
Kamesh Munagala and Kangning Wang.
\newblock Improved metric distortion for deterministic social choice rules.
\newblock In \emph{Proceedings of the 2019 {ACM} Conference on Economics and
  Computation ({EC})}, pages 245--262, 2019.

\bibitem[Nisan et~al.(2007)]{nisan2007introduction}
Noam Nisan et~al.
\newblock Introduction to mechanism design (for computer scientists).
\newblock \emph{Algorithmic Game Theory}, 9:\penalty0 209--242, 2007.

\bibitem[Pierczynski and Skowron(2019)]{pierczynski2019approval}
Grzegorz Pierczynski and Piotr Skowron.
\newblock Approval-based elections and distortion of voting rules.
\newblock In \emph{Proceedings of the 28th International Joint Conference on
  Artificial Intelligence ({IJCAI})}, pages 543--549, 2019.

\bibitem[Procaccia and Rosenschein(2006)]{procaccia2006distortion}
Ariel~D. Procaccia and Jeffrey~S. Rosenschein.
\newblock The distortion of cardinal preferences in voting.
\newblock In \emph{International Workshop on Cooperative Information Agents
  ({CIA})}, pages 317--331, 2006.

\bibitem[Sen(1986)]{sen1986social}
Amartya Sen.
\newblock Social choice theory.
\newblock \emph{Handbook of mathematical economics}, 3:\penalty0 1073--1181,
  1986.

\bibitem[Skowron and Elkind(2017)]{skowron2017stv}
Piotr~K. Skowron and Edith Elkind.
\newblock Social choice under metric preferences: Scoring rules and {STV}.
\newblock In \emph{Proceedings of the Thirty-First {AAAI} Conference on
  Artificial Intelligence ({AAAI})}, pages 706--712, 2017.

\bibitem[Tideman(1987)]{tideman1987independence}
T.~Nicolaus Tideman.
\newblock Independence of clones as a criterion for voting rules.
\newblock \emph{Social Choice and Welfare}, 4\penalty0 (3):\penalty0 185--206,
  1987.

\bibitem[Vazirani(2013)]{vazirani2013approximation}
Vijay~V. Vazirani.
\newblock \emph{Approximation algorithms}.
\newblock Springer Science \& Business Media, 2013.

\bibitem[Von~Neumann and Morgenstern(1953)]{morgenstern1953theory}
John Von~Neumann and Oskar Morgenstern.
\newblock \emph{Theory of games and economic behavior}.
\newblock Princeton university press, 1953.

\end{thebibliography}

\end{document}